\documentclass[aps,prl,amsmath,amssymb,reprint,superscriptaddress,twocolumn,showpacs]{revtex4-2}

\usepackage{graphicx}
\usepackage{amsmath}
\usepackage{amsfonts}
\usepackage{amssymb}
\usepackage{pstricks}
\usepackage{psfrag}
\usepackage{epsfig}
\usepackage{changes}
\usepackage[ansinew]{inputenc}
\usepackage{mathptmx}
\usepackage[normalem]{ulem}
\usepackage{textcomp}
\usepackage{ulem}
\usepackage{bm}

\begin{document}

\title{Emergent Synchronization and Defect Dynamics in Confined Chiral Active Suspensions}

\author{Zaiyi Shen}
\email[]{zaiyi.shen@pku.edu.cn}
\affiliation{State Key Laboratory for Turbulence and Complex Systems, School of Mechanics and Engineering Science, Peking University, Beijing 100871, China}
\author{Leilei Wang}
\affiliation{State Key Laboratory of Nonlinear Mechanics, Beijing Key Laboratory of Engineered Construction and Mechanobiology, Institute of Mechanics, Chinese Academy of Sciences, Beijing 100190, China}
\author{Shishuang Zhang}
\affiliation{State Key Laboratory of Nonlinear Mechanics, Beijing Key Laboratory of Engineered Construction and Mechanobiology, Institute of Mechanics, Chinese Academy of Sciences, Beijing 100190, China}
\affiliation{School of Building Services Science and Engineering, Xi'an University of Architecture and Technology, Xi'an 710055, China}
\author{Chenlu Li}
\affiliation{State Key Laboratory of Nonlinear Mechanics, Beijing Key Laboratory of Engineered Construction and Mechanobiology, Institute of Mechanics, Chinese Academy of Sciences, Beijing 100190, China}
\affiliation{School of Building Services Science and Engineering, Xi'an University of Architecture and Technology, Xi'an 710055, China}
\author{Kaili Xie}
\affiliation{Van der Waals-Zeeman Institute, Institute of Physics, University of Amsterdam, 1098XH, Amsterdam, The Netherlands}
\author{Xu Zheng}
\email[]{zhengxu@lnm.imech.ac.cn}
\affiliation{State Key Laboratory of Nonlinear Mechanics, Beijing Key Laboratory of Engineered Construction and Mechanobiology, Institute of Mechanics, Chinese Academy of Sciences, Beijing 100190, China}
\author{Juho S. Lintuvuori}
\email[]{juho.lintuvuori@u-bordeaux.fr}
\affiliation{Univ. Bordeaux, CNRS, LOMA, UMR 5798, F-33400 Talence, France}

\date{\today}

\begin{abstract}
Hydrodynamic interactions can generate rich emergent structures in active matter systems. Using large-scale hydrodynamic simulations, we demonstrate that hydrodynamic coupling alone can drive spontaneous self-organization across a hierarchy of spatial and temporal scales in confined suspensions of torque-driven particles at moderate Reynolds numbers. Spinners first self-assemble into dimers, which crystallize into a hexatic lattice and subsequently undergo a collective tilting instability. The resulting tilted dimers rotate and synchronize through hydrodynamic repulsion, which can be tuned by the Reynolds number. Upon synchronization, the polar director develops splay and bend deformations and nucleates topological defects with charges of $\pm1$. These defects induce long-wavelength concentration gradients and drive crystal vortex dynamics spanning hundreds of particle diameters. Our results reveal a purely hydrodynamic route to synchronization and defect-mediated dynamics in chiral active matter, without explicit alignment rules or interparticle forces.
\end{abstract}


\maketitle



Active matter systems, composed of energy-consuming units, can exhibit a wealth of emergent behaviors~\cite{marchetti2013hydrodynamics,bechinger2016active}, including flocking~\cite{bricard2013emergence}, turbulence~\cite{alert2022active}, and motility-induced phase separation~\cite{cates2015motility}. A particularly rich class involves chiral active matter, where rotational constituents can break mirror and time-reversal symmetries~\cite{o2022time}, enabling circulating flows~\cite{van2016spatiotemporal,liu2020oscillating,li2024robust}, nonreciprocal interactions~\cite{poncet2022soft,bililign2022motile}, and unconventional transport~\cite{poggioli2023odd,hargus2025flux}. Recent studies with magnetically driven rotors have revealed edge currents and odd viscosity in spinner materials~\cite{soni2019odd}, self-healing lattices at fluid interfaces~\cite{han2020reconfigurable}, and directional exchange under confinement~\cite{massana2020emergent,meng2020field}. In biological systems, hydrodynamic coupling alone has been shown to generate odd dynamics among starfish embryos~\cite{tan2022odd} and crystallization of spinning bacteria~\cite{petroff2015fast}, while theoretical models predict rotor synchronization~\cite{uchida2010synchronization} via fluid-mediated interactions. Yet most of these investigations either focus on the zero-inertia regime or rely on explicit alignment rules or external fields, leaving open whether hydrodynamic flows alone can organize structure at finite Reynolds numbers. 

At moderate Reynolds numbers, rotating particles have been observed to form vortex condensates~\cite{shen2020hydrodynamic}, two-phase crystals~\cite{shen2020two}, and other emergent structures~\cite{grzybowski2000dynamic,goto2015purely,shen2023collective}. Despite growing recognition of fluid inertia as a key driver of self-organization, its full implications remain largely unexplored and may hold the key to revealing a much broader spectrum of emergent behaviors in active matter.

\begin{figure*}
\centering
\includegraphics[width=2\columnwidth]{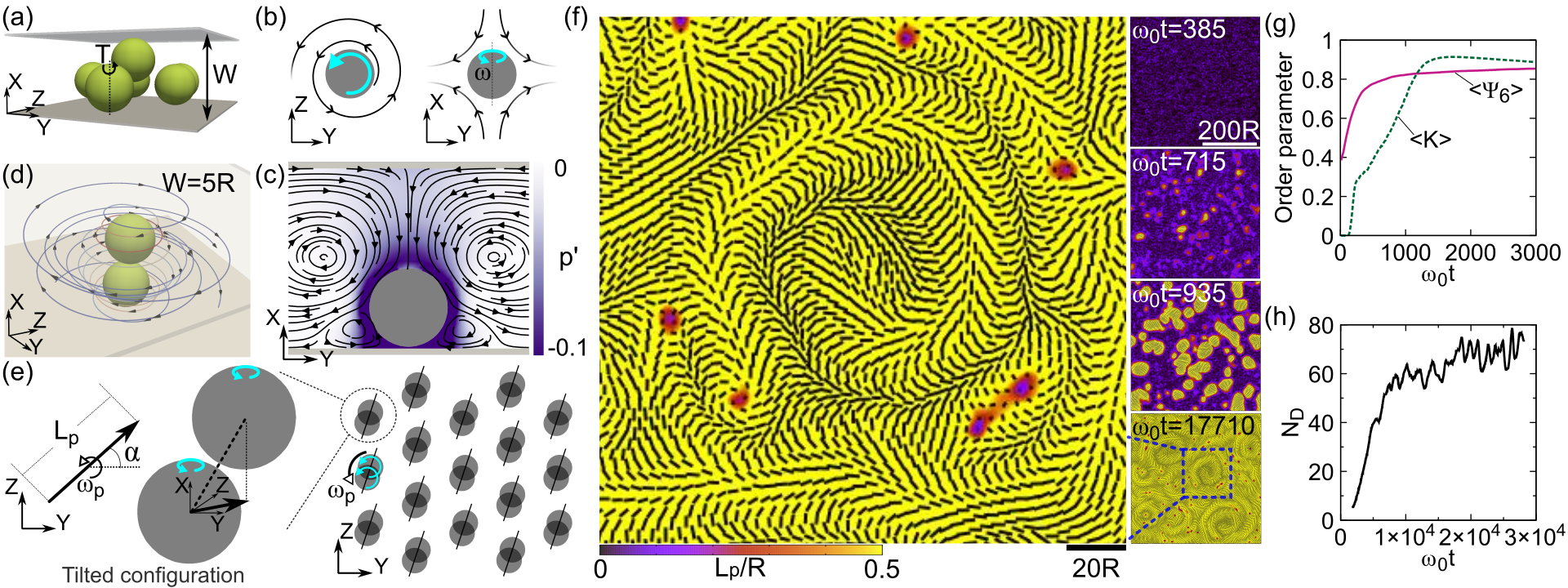}
\caption{\footnotesize {{\bf Multiscale self-organization of confined spinners.} 
(a) Schematic of the system. 
(b) Schematic representation of the streamlines generated by a spinner at $\mathrm{Re} \sim 10$. 
(c) Simulated flow field of an isolated spinner at $\mathrm{Re} \approx 7$ confined between walls separated by $W \approx 5R$. A recirculating flow arises due to the confinement, as indicated by the streamlines. The color map shows the pressure field, with low-pressure regions appearing near the spinner surface. 
(d) Example of spinner pairing and the surrounding streamlines of a stable vertical dimer under confinement ($W \approx 5R$). 
(e) Schematic illustrating the crystallization of tilted dimers and rotational synchronization. Black lines (magnified $\times 4$) represent the alignment $\mathbf{L}_p$, calculated as the projection of the vector connecting the centers of two paired spinners onto the plane perpendicular to the spinning axis. 
(f) Right: Time series of snapshots showing the evolution of tilted dimer configurations. At early times, the dimers are vertical ($L_p\approx 0$). Spontaneous tilting ($L_p > 0$) emerges and spreads across the system. The color map indicates the spatial distribution of $L_p$. Left: Zoom-in view of the boxed area in the right panel. Black lines show the alignment of the dimers given by $\mathbf{L}_p$} (magnified $\times 4$). Darker regions highlight defect locations, where vertical pairing persists ($L_p\approx 0$). Simulation corresponds to $N=30720$ spinners at $\mathrm{Re} \approx 10$, confined within a computational domain of $5R\times 480R\times 480R$. 
(g) Time evolution of the hexatic order parameter $\Psi_6$ and the Kuramoto order parameter $K$. 
(h) Time evolution of the total number of defects $N_D$.
\label{fig1}}
\end{figure*}

Here, we demonstrate that weakly inertial hydrodynamics can serve as a standalone organizing principle in confined suspensions of torque-driven particles. Using large-scale simulations, we uncover a cascade of emergent structures: rotating dimers, hexatic crystals, synchronized dimer rotation, active topological defects and ultimately crystal vortex dynamics governed by defects-induced concentration gradients, spanning progressively larger spatial and temporal scales. Each of these states arises solely from fluid-mediated interactions.

We study the collective dynamics of spherical spinners confined between two flat walls [Fig.~\ref{fig1} (a)], using lattice Boltzmann simulations~\cite{kevinstratford_2025_15358003}. The particles, density-matched with the surrounding fluid, are subjected to a constant torque $T$ around the axis perpendicular to the confining walls [Fig.~\ref{fig1} (a)], resulting in steady spinning with a frequency $\omega$. Periodic boundary conditions are applied along the directions parallel to the walls. Hydrodynamic interactions are fully resolved by solving the Navier-Stokes equations, with non-slip conditions enforced via the bounce-back method ~\cite{ladd1994numerical1}. 
The particle Reynolds number is defined as $\mathrm{Re} = \rho \omega_0 R^2/\mu$,
where $\omega_0 = T/8\pi\mu R^3$ is the Stokes-limit spinning frequency, with $R$, $\rho$ and $\mu$ denoting the particle radius, fluid density, and viscosity, respectively. The Reynolds number serves as the primary control parameter.
A short-range repulsive potential is implemented to prevent solid surface overlaps, including particle-particle and particle-wall interactions ~\cite{shen2018hydrodynamic}.

\noindent \textbf{Multiscale self-organization} -- Starting from randomly distributed spinners [Fig.~\ref{fig1} (a)], we observe a remarkable hierarchy of self-organized states  arising solely from hydrodynamic interactions at $\mathrm{Re} \sim 10$. 

A single spinner generates a secondary flow [Fig.~\ref{fig1} (b)] consisting of a radial component that advects fluid inward near the poles and outward along the equatorial plane ~\cite{bickley1938lxv}, while the azimuthal component maintains axisymmetric ~\cite{climent2007dynamic,shen2020hydrodynamic}. 
Confinement breaks the pole symmetry, leading to an effective attraction along the spinning axis~\cite{liu2010wall}.  When placed between two parallel walls, the secondary flow field of a single spinner develops a recirculating pattern, pumping fluid from the equator toward the pole farther from the nearby wall [Fig.~\ref{fig1} (c)]. Based on these flow patterns, we hypothesize that two spinners at opposite walls would experience hydrodynamic attraction and form a bound state [Fig.~\ref{fig1} (d)]. Simulations of two randomly positioned spinners confirm this: the particles rapidly trap one another and orbit around a common center, eventually stabilizing into a steady configuration [Fig.~\ref{fig1} (d) and Movie 1]. The final structure of the dimer depends on the wall separation $W$ (Movie 1). For a moderate confinement ($W \approx 5R$), sufficient vertical space allows the spinners to align vertically with negligible lateral displacement [Fig.~\ref{fig1} (d)]. 

At larger scales, the dimers spontaneously self-organize into a hexatic crystal [Fig.~\ref{fig1} (e)]. At zero Reynolds number, it is known that a combination of steric repulsion and mixing driven by azimuthal (rotational) flows promotes the rapid crystallization of spinners~\cite{oppenheimer2019rotating}. Here, at a finite Reynolds number, the repulsion emerges naturally from the secondary flow generated by each spinner [Fig.~\ref{fig1} (b)], leading to a similarly rapid hexatic ordering of the dimers.

At sufficiently high concentrations, hydrodynamic interactions can trigger a collective tilting instability. Considering an initially vertical configuration, local perturbations in spinner positions grow and lead to spontaneous tilting that propagates throughout the system (Movie 2). This instability drives the dimers to adopt tilted configurations characterized by an orientation field $\mathbf{L}_p$ corresponding to the projection of the center-to-center separation of the two spinners onto the plane of rotation [left panel in Fig.~\ref{fig1} (e)]. Each spinner orbits its partner around a shared center of mass with a characteristic angular speed $\omega_p < \omega_0$, while the overall crystalline order of the dimer centers is maintained [Fig.~\ref{fig1} (e)]. The rotation of dimers tends to synchronize with their neighbors, and at larger scales, this local phase locking gives rise to spiral structures in the orientation field $\mathbf{L}_p$, punctuated by the nucleation of topological defects [Fig.~\ref{fig1} (f)]. 

We demonstrate the formation of these dynamic states in a large-scale simulation of $N = 30720$ spinners (volume fraction $\phi\approx 11$\% ) confined between parallel walls separated by $W = 5R$ at a Reynolds number $\mathrm{Re} \approx 10$ [Fig.~\ref{fig1} (f-h)]. Randomly initialized spinner dimers rapidly self-organize into a hexatic crystal, as quantified by the local hexatic order parameter $\displaystyle \Psi_6 = \left| \frac{1}{n_j} \sum_k \exp(i6\theta_{jk}) \right|$, where $n_j$ is the number of nearest neighbors of the dimer $j$, and $\theta_{jk}$ is the angle between the vector connecting the dimers $j$ and $k$ and a fixed horizontal axis. The spatial average $\langle \Psi_6 \rangle$ quickly rises to 0.85 [Fig.~\ref{fig1} (g)], indicating rapid crystallization after $\omega_0 t\approx 500$, driven by hydrodynamic repulsion. 

At the initial stage after rapid crystallization, the particle pairs are vertical ($L_b\sim 0$) [see {\it e.g.} $\omega_0 t\approx 385$ in the insets of Fig.~\ref{fig1} (f)]. The hydrodynamic interactions gradually perturb the vertical pairing, inducing tilting ($L_b > 0$) and consequently dimer rotation that propagates through the system [$\omega_0 t\approx 715$ and $\omega_0 t\approx 935$ in the insets of Fig.~\ref{fig1} (f) and Movie 3]. This propagation is subsequently followed by spontaneous rotational synchronization, corresponding to the alignment of the tilted dimers [Fig.~\ref{fig1} (f) and Movie 3].

To evaluate  the synchronization in the system, we calculate averaged local Kuramoto order parameter $\displaystyle \langle K \rangle = \langle \frac{1}{m_j}  \left| \sum_k \exp(i\alpha_k)\right| \rangle$, where $\alpha_k$ is the in-plane orientation angle of the dimer $k$ (left panel in Fig.~\ref{fig1}e) within a circular region of radius $10R$ centered on the dimer $j$ and $m_j$ is the total number of dimers within the region. The value of $\langle K \rangle$ rises steadily and reaches approximately 0.9 after $\omega_0 t \approx 1200$ [Fig.\ref{fig1}(g)], indicating a high degree of local rotational synchronization.

\begin{figure}
\centering
\includegraphics[width=1\columnwidth]{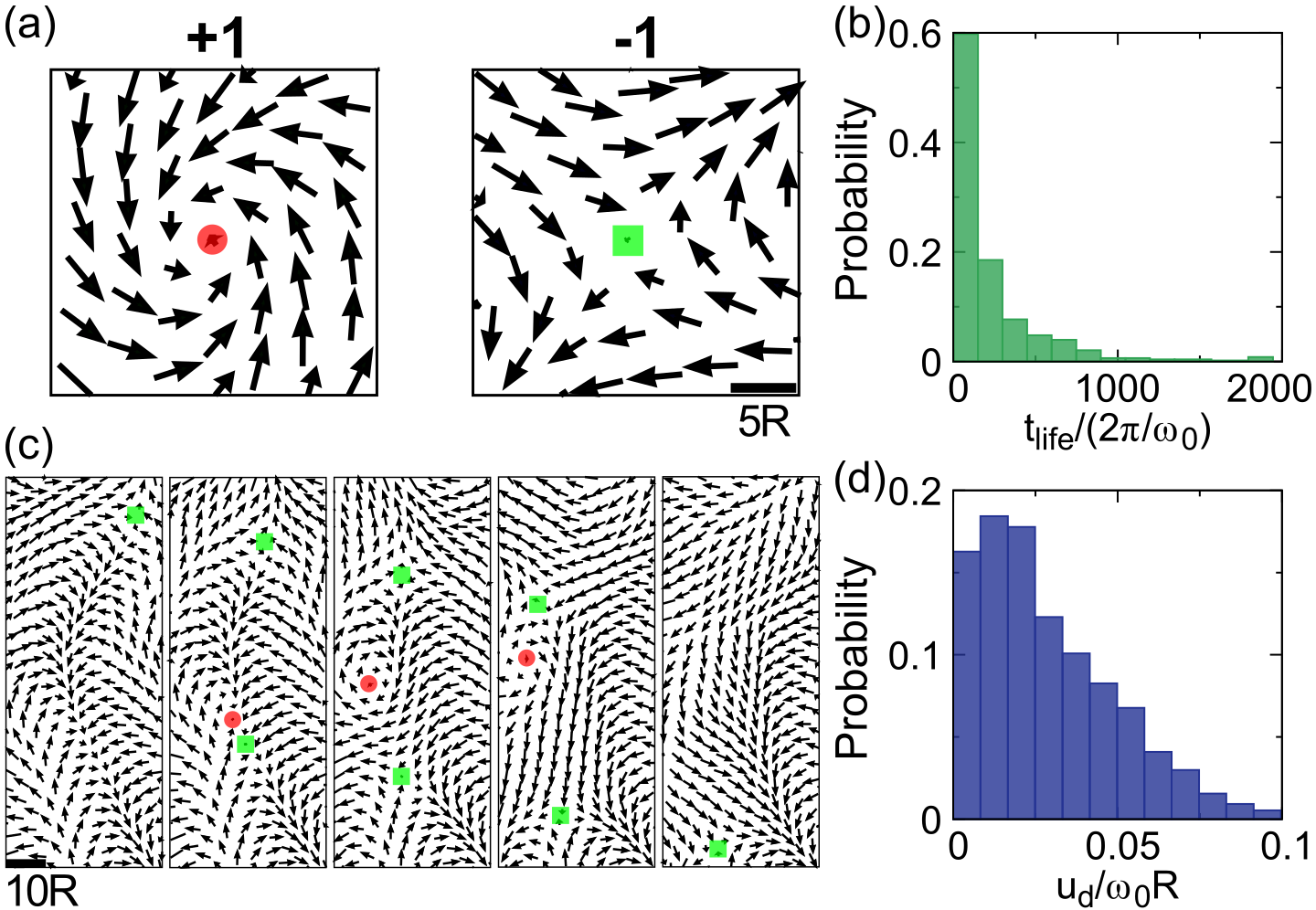}
\caption{\footnotesize {\bf Active defects.} 
(a) Instantaneous orientation structures around a $+1/-1$ defect.
(b) Probability distribution of defect lifetimes.
(c) Example of a short-lived $+1/-1$ defect pair undergoing spontaneous nucleation, movement, and annihilation.
(d) Probability distribution of defect speeds. 
\label{fig2}}
\end{figure}

At large scales, global synchronization of dimer rotations is not attained. Instead, the system self-organizes into multiple locally synchronized domains separated by finite phase lags. To accommodate smooth phase variation between these regions, the orientation field develops bend and splay deformations and formation of active defects. The defects can be identified as regions where the bond length $L_b$ is significantly reduced  [dark red regions in Fig.\ref{fig1} (f), corresponding to $L_b\sim 0$]. In the bend and splay zones, $L_b$ remains finite [yellow areas in Fig.~\ref{fig1} (f)]. By tracking the spatial distribution of $L_b$, we estimate the total number of defects $N_D$, which fluctuates between approximately 60 and 80 in the steady state [Fig.~\ref{fig1} (h)]. 

The deformation and defects are reminiscent of those found in nematic liquid crystals. However, unlike conventional nematics, the dimers rotate persistently, and the orientation field $\hat{\mathbf{n}} = \cos\alpha \hat{\mathbf{y}} + \sin\alpha \hat{\mathbf{z}}$ is polar rather than headless [Fig.~\ref{fig2} (a)]. The formation of topological defects with charges of $\pm1$ is observed [Fig.~\ref{fig2} (a)]. In the steady state, most defects exhibit lifetimes $t_{\text{life}} < 100 \cdot 2\pi/\omega_0$ [Fig.~\ref{fig2} (b)] and undergo continuously migration, annihilation, or nucleation through local reconfigurations of the orientation field [Fig.~\ref{fig2} (c) and Movie 4]. Their migration velocity ($\sim 0.01 \omega_0 R$) is significantly slower than the tangential surface velocity of the spinners [Fig.~\ref{fig2} (d)]. Due to the periodic boundary conditions, topologically equivalent to a torus, the total topological charge remains conserved at $q = 0$ throughout the dynamics.

\begin{figure}
\centering
\includegraphics[width=1\columnwidth]{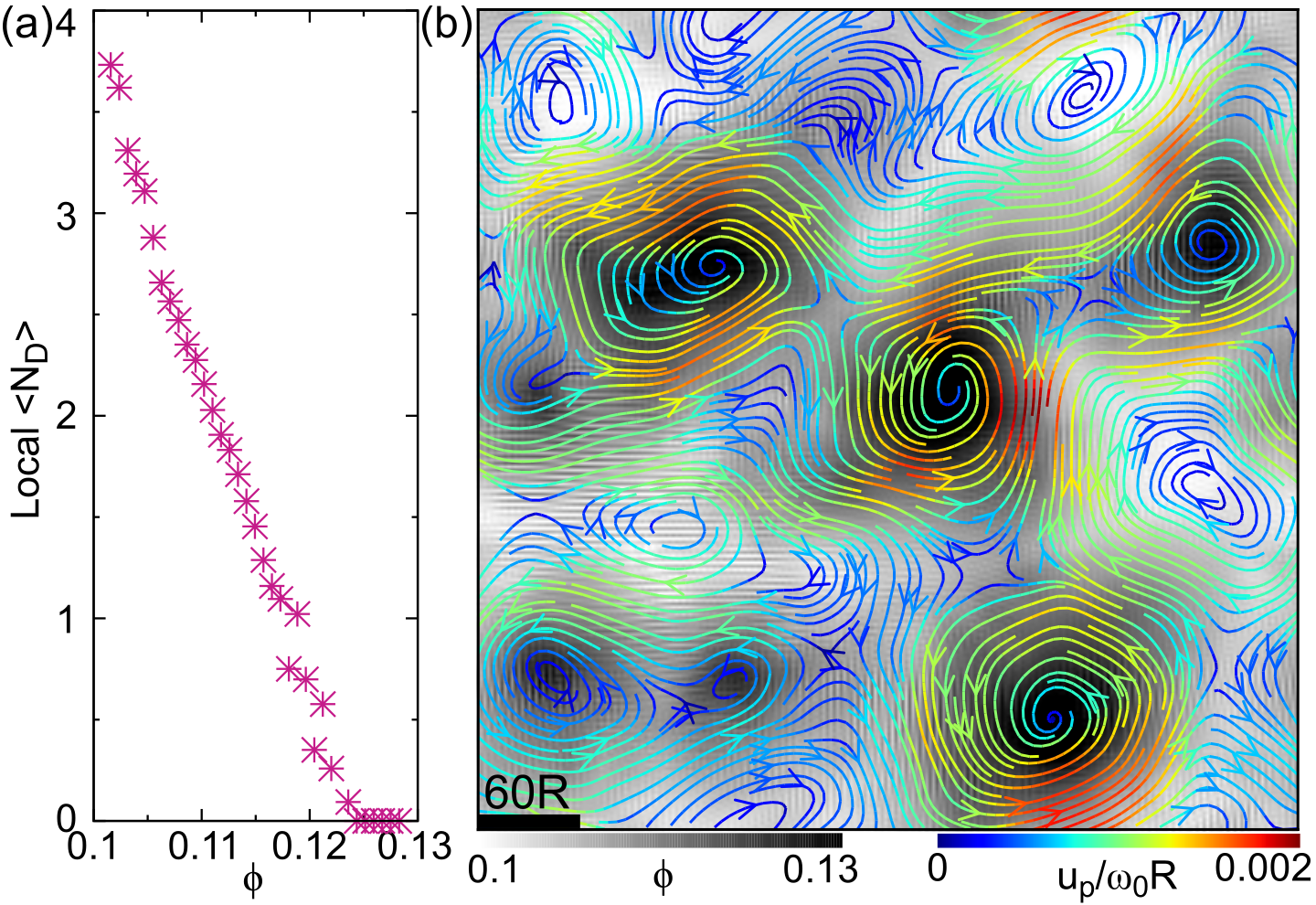}
\caption{\footnotesize {\bf Vortex dynamics.} 
(a) The average number of defects $N_D$ decreases with increasing volume fraction of a local domain $\phi$. 
(b) Regions with high (black) or low (white) local volume fractions exhibit slow vortical motion. The black-and-white shading indicates the local volume fraction. The motion of the dimer centers is represented by streamlines, color coded by the observed speed. \label{fig3}}
\end{figure}

Due to the presence of topological defects, the long-time and large-scale dynamics of the system become dominated by defect-driven processes. The defects not only shape the orientation field, but also induce concentration gradients in the particle distribution. Around $\pm1$ defects, local orientational disorder increases the effective repulsive potential between the dimers, leading to a reduction in the local packing density. Indeed, we observe a strong linear relationship between the number of local defects and the local particle concentration [Fig.~\ref{fig3} (a)]. Regions with more defects exhibit lower dimer density .

As a result, the dimer crystal develops long-wavelength density modulations [Fig.~\ref{fig3} (b)], where regions of enhanced and depleted concentrations rotate collectively on timescales far slower than the intrinsic spinner frequency. Intriguingly, the concentrated regions rotate counterclockwise, in the same direction as the individual spinners, while depleted regions exhibit counter-rotating vortices [Fig.~\ref{fig3} (b)]. These emergent vortex structures span hundreds of particle diameters and represent the largest dynamic scale in the system. Despite the continuous nucleation and annihilation of short-lived defects, the global vortex lattice remains stable within the duration of the simulations, highlighting the robustness of the emergent multiscale order.

\noindent \textbf{Hydrodynamic mechanism for synchronization} -- The local alignment of the dimers is the key ingredient behind the observed multi-scale dynamics. To isolate the mechanism leading to synchronization, we investigate a slightly stronger confinement ($W = 4R$), where the geometry imposes tilted configuration, by fixing the center-to-center distance within each dimer (Movie 1). This setup yields a system of dimers with fixed tilt angles, allowing us to focus on their in-plane rotational dynamics. 

\begin{figure}
\centering
\includegraphics[width=1\columnwidth]{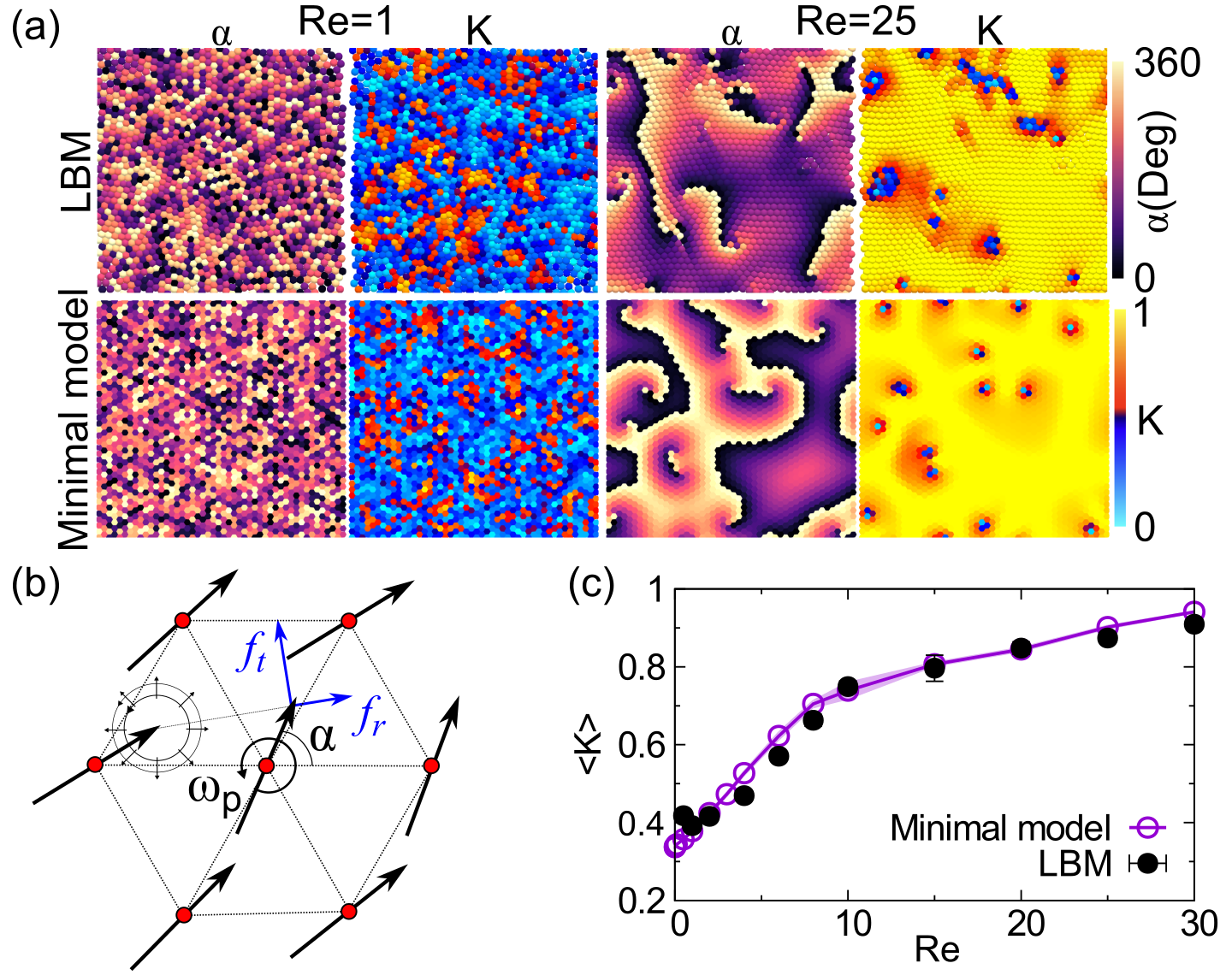}
\caption{\footnotesize {\bf Rotational synchronization driven by hydrodynamic interactions.} 
(a) Spatial fields of dimer orientation angle $\alpha$ and local synchronization parameter $K$. 
(b) Schematic of the minimal model describing rotational dynamics of dimers arranged on a hexatic lattice. Each dimer is represented by a rotating vector, and hydrodynamic interactions are modeled via transverse $f_t$ and repulsive $f_r$ forces. 
(c) Mean Kuramoto order parameter $\langle K \rangle$ as a function of Reynolds number $\mathrm{Re}$. Results from both hydrodynamic simulations and the minimal model show a consistent increase in synchronization with $\mathrm{Re}$. 
\label{fig4}}
\end{figure}

In simulations of $N = 3200$ spinners confined in the domain of $4R \times 240R \times 240R$ (volume fraction $\phi \approx 5.8\%$), we investigate the orientational ordering and synchronization dynamics of dimers with varying Reynolds numbers. At moderate Reynolds number ($\mathrm{Re} \approx 25$), the orientation field $\hat{\mathbf{n}}(\mathbf{r})$ exhibits extended nematic domains with well-defined alignment, accompanied by high local synchronization measured by the Kuramoto order parameter ($K \approx 1$) [Fig.~\ref{fig4} (a)]. In contrast, at low Reynolds number ($\mathrm{Re} \approx 1$), the orientation field becomes isotropic and disordered, indicating the absence of synchronization [Fig.~\ref{fig4} (a)].

Hydrodynamic interactions in the system are highly complex due to the presence of many particles and the confinement. The flow field around a single spinner features both radial ($v_r$) and tangential ($v_t$) components in the equatorial plane [Fig.~\ref{fig1} (b)]. We hypothesize that the dominant interactions arise from these components, which generate effective in-plane radial repulsion and transverse forces between neighboring spinners. From simulations of two interacting spinners near a wall under the same confinement ($W = 4R$), we extract the force scaling as $f_t = 4.56 F_0 (r/R)^{-3.85}$ and $f_r = 0.08 F_0 \mathrm{Re} (r/R)^{-2.9}$, where $F_0=6\pi\mu\omega_0 R^2$ (see SM).

To isolate the role of these interactions in synchronization, we construct a minimal model in which rotating dimers are fixed on hexagonal lattice sites [Fig.~\ref{fig4} (b)]. Each dimer rotates with a natural angular velocity $\mathbf{\omega}_d$ and experiences both radial and transverse forces from its neighbors. Assuming an overdamped regime, the orientation of the $i$th dimer evolves according to
\begin{equation}
\bm{\omega}_{p}^{i} = \frac{\mathrm{d} \bm{\alpha}_i}{\mathrm{d} t} = \bm{\omega}_d + \sum_{j} \frac{\bm{f}_{ij} \times \bm{\hat{n}}_i}{\xi a},
\label{eq1}
\end{equation}
where $\bm{f}_{ij}$ is the hydrodynamic force exerted by the $j$th dimer on the $i$th, decomposed into transverse ($f_t \bm{\hat{t}}_{ij}$) and radial ($f_r \bm{\hat{r}}_{ij}$) components. Here, $\bm{\hat{n}}_i$ is the director of the $i$th dimer, $a$ is the distance from the lattice site to the spinner within a dimer, and $\xi$ is the viscous drag coefficient, given by Stokes drag $\xi = 6\pi \mu R$.

Using the lattice distance $d\approx 6.4R$ (corresponding to the volume fraction $\phi \approx 5.8\%$ in hydrodynamic simulations), this minimal model captures the essential features of the full hydrodynamic simulations: it exhibits large-scale nematic domains and strong local synchronization at $\mathrm{Re} \approx 25$, and transitions to an isotropic, disordered state at $\mathrm{Re} \approx 1$ [Fig.~\ref{fig4} (a)], in excellent agreement with the full simulations [Fig.~\ref{fig4} (c)].

The ratio between the radial and transverse interactions scales linearly with the Reynolds number $f_r/f_t \sim \mathrm{Re}$. Comparison across simulations reveals that radial repulsion promotes phase locking and synchronization at higher $\mathrm{Re}$, while transverse forces induce phase divergence and disorder at lower $\mathrm{Re}$. To test the robustness of this mechanism, we conducted additional simulations at different particle concentrations. In all cases, both the hydrodynamic simulations and the minimal model exhibit a clear transition from a disordered state to synchronized rotation as $\mathrm{Re}$ increases (see SM), confirming the consistency of the results. 

Furthermore, to examine the generality of this behavior, we formulated generalized interaction forms unrelated to fluid flow $f_t=\xi \omega R r^{-3}$ and $f_r=\beta \xi \omega R r^{-3}$, where $\beta=f_r/f_t$ defines the ratio between repulsive and transverse forces.  The numerics demonstrate that purely radial interactions ($f_r/f_t \to +\infty$) consistently stabilize synchronization, whereas purely transverse interactions ($f_r/f_t \to 0$) destabilize it (for more details see SM).

Together, these findings reveal that the emergence of collective synchronization in dimer rotation is governed by the competition between radial and transverse interactions. Remarkably, this balance can be tuned using a single control parameter, the Reynolds number $\mathrm{Re}$, providing a natural mechanism to switch on and off the collective phase ordering.

\noindent \textbf{Conclusion} -- Our study demonstrates that hydrodynamic interactions alone, even at moderate Reynolds numbers, can drive a cascade of self-organization across multiple spatial and temporal scales in confined chiral active matter. Starting from torque-driven spinners, the system spontaneously develops a sequence of emergent states: stable rotating dimers, hexatic crystals, active chiral polar nematics, topological defects, and ultimately large-scale crystal vortex dynamics -- all without any explicit alignment rules or external fields.

These findings establish fluid-mediated interactions as a powerful, self-sufficient organizing principle in active systems, capable of coordinating collective dynamics from microscopic to macroscopic scales. Beyond advancing our understanding of pattern formation in inertial chiral fluids, our results suggest new strategies for engineering tunable multiscale order in synthetic active materials, microrobotic swarms, and biologically inspired systems. More broadly, they reveal a general framework in which nonequilibrium structure emerges naturally from the interplay of chirality, inertia, and confinement -- offering routes to controlling active matter through hydrodynamic interactions.

\begin{acknowledgments}
ZS acknowledge the Natural Science Foundation of Beijing, China through Grant No. 1252020 for funding; ZS and JSL acknowledge IdEx (Initiative d'Excellence) Bordeaux and the  French  National  Research  Agency  through  Contract No. ANR-19-CE06-0012 for funding, Curta cluster for computational time. XZ and LW acknowledge the National Key R$\&$D Program of China (2022YFF0503504), the Strategic Priority Research Program of Chinese Academy of Sciences (Grant no. XDB0620102, XDA0470203), National Natural Science Foundation of China (Grant No. 12302357, 12472273).
\end{acknowledgments}

\bibliography{ref}



\end{document}